\documentclass[runningheads]{llncs}
\usepackage{amssymb}
\usepackage{graphicx}
\usepackage{cite}
\usepackage{color}
\usepackage{textcomp}
\usepackage{url}
\usepackage{caption}
\usepackage{siunitx}

\usepackage{academicons}
\usepackage{hyperref}
\usepackage{xcolor}
\newcommand{\orcid}[1]{\href{https://orcid.org/#1}{\includegraphics[scale=0.08]{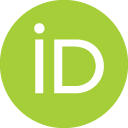}}}

\usepackage{url}
\urldef{\mailsa}\path|{alfred.hofmann, ursula.barth, ingrid.haas, frank.holzwarth,|
\urldef{\mailsb}\path|anna.kramer, leonie.kunz, christine.reiss, nicole.sator,|
\urldef{\mailsc}\path|erika.siebert-cole, peter.strasser, lncs}@springer.com|    
\newcommand{\keywords}[1]{\par\addvspace\baselineskip
\noindent\keywordname\enspace\ignorespaces#1}

\begin{document}

\mainmatter  % start of an individual contribution

\title{A Computational Study of the Reaction Cyanoacetylene and Cyano Radical leading to 2-Butynedinitrile and Hydrogen Radical }

\titlerunning{A Computational Study of the Reaction HC$_{3}$N + CN Leading to C$_{4}$N$_{2}$ + H}
%}
\author{Emília Valença Ferreira de Aragão\inst{1,2}\orcid{0000-0002-8067-0914} 
\and Noelia Faginas-Lago\inst{1}\orcid{0000-0002-4056-3364} 
\and Marzio Rosi\inst{3}\orcid{0000-0002-1264-3877} 
\and Luca Mancini\inst{1}
\and  Nadia Balucani\inst{1}\orcid{0000-0001-5121-5683}
\and Dimitrios Skouteris\inst{2}\\\
}

\authorrunning{E.V.F. Aragão et al.}

\institute{
Dipartimento di Chimica, Biologia e Biotecnologie,\\ Universit\`{a} degli Studi di Perugia, 06123 Perugia, Italy\\ 
\email{\{emilia.dearagao,luca.mancini2\}@studenti.unipg.it}\\
\email{\{noelia.faginaslago,nadia.balucani\}@unipg.it}\\
\and 
Master-up srl, Via Sicilia 41, 06128, Italy\\
\email{\{emilia.dearagao,d.skouteris\}@master-up.it}\\
\and
 Dipartimento di Ingegneria Civile ed Ambientale,\\ Universit\`{a} degli Studi di Perugia, 06125 Perugia, Italy\\
 \email{marzio.rosi@unipg.it}
}

\toctitle{Lecture Notes in Computer Science}
\tocauthor{Authors' Instructions}
\maketitle

\begin{abstract}

The present work focuses on the characterization of the reaction between cyanoacetylene and cyano radical by electronic structure calculations of the stationary points along the minimum energy path.
One channel, leading to C$_{4}$N$_{2}$ (2-Butynedinitrile) + H, was selected due to the importance of its products.
Using different ab initio methods, a number of stationary points of the potential energy surface were characterized. The energy values of these minima were compared in order to weight the computational costs in relation to chemical accuracy.
The results of this works suggests that B2PLYP (and B2PLYPD3) gave a better description of the saddle point geometry, while B3LYP works better for minima.

\end{abstract}

\section{Introduction}
Cyanopolyynes are a family of carbon-chain molecules that have been detected in numerous objects of the Interstellar medium (ISM), such as hot cores, star forming regions and cold clouds~\cite{ wyrowski1999vibrationally, taniguchi2018survey, mendoza2018search, takano1998observations}.
They are all linear molecules, with alternating carbon-carbon triple and single bonds. The simplest cyanopolyyne, HC$_{3}$N, has been among the first organic molecules to be detected in the ISM \cite{turner1971detection} and  
up to date also HC$_{5}$N, HC$_{7}$N, HC$_{9}$N and HC$_{11}$N have been detected at least once in the ISM~\cite{broten1978Adetection,bell1997detection} (the detection of HC$_{11}$N, however, has been recently disputed by Loomis et al. \cite{loomis2016non} and Cordiner et al. \cite{cordiner2017deep}).
HC$_{3}$N and HC$_{5}$N are also abundant in solar-type protostars (see for instance a recent work on IRAS 16293-2422 by Jaber Al-Edhari et al. \cite{jaber2017history}).\\
\indent
The shortest and most abundant member of the cyanopolyyne family, HC$_{3}$N (cyanoacetylene), is a precursor of chain elongation reactions: the successive addition of C$_{2}$H molecules generates the other members of its family, as summarised by Cheikh et al~\cite{cheikh2013low}. According to the same authors, however, addition of CN radical instead of C$_{2}$H would result in a chain termination reaction by the formation of dicyanopolyynes, NC(CC)$_{n}$CN. These species have not been observed in the ISM so far because they lack a permanent electric dipole moment and cannot be detected through their rotational spectrum. However, it has been suggested that they are abundant in interstellar and circumstellar clouds \cite{petrie2003nccn}. The recent detection of NCCN \cite{agundez2015probing} via its protonated form NCCNH$^+$ seems to corroborate the suggestion by Petrie et al. \cite{petrie2003nccn}\\
\indent
In 2004, Petrie and Osamura had explored through computational means the formation of C$_{4}$N$_{2}$ (2-Butynedinitrile) in Titan's atmosphere through many pathways~\cite{petrie2004nccn}. 
They had found in particular that the cyano radical addition to cyanoacetylene leads to C$_{4}$N$_{2}$.
In order to characterize all the stationary points of the potential energy surface, the authors had used at the time the hybrid density functional method B3LYP in conjunction with triple-split valence Gaussian basis set 6-311G** for geometry optimizations and vibrational frequency calculations. 
Moreover, they had also performed single-point calculations with CCSD(T) in conjunction with aug-cc-pVDZ basis-set, a choice made at the time due to computational costs.
As computational power has risen in the last 15 years and new methods have been implemented into quantum chemistry software, more accurate results for the geometries and energies can be obtained. 
\indent
In our laboratory we have already investigated several reactions of astrochemical interest providing insightful results for the understanding of processes observed in the ISM~\cite{podio2017silicon,skouteris2019interstellar} including those involving or leading N-bearing organic molecules \cite{skouteris2015dimerization,balucaniMolAp2018,sleimanPCCP2018}.
Recently, we have focused on studying the reaction between HC$_{3}$N and CN in collaboration with the experimental part of the Perugia group. 
The preliminary investigation of the potential energy surface of the system HC$_{3}$N + CN showed the presence of a number of reactive channels. The search for intermediate and product species always involves the computation of four carbon atoms, two nitrogen atoms and one hydrogen atom, i.e. 39 electrons in total.
Ab initio calculations can become prohibitively expensive with a rising number of electrons, therefore a balance between chemical accuracy and computational cost must be reached.
The focus of this paper is the exit channel of the reaction between cyanoacetylene and cyano radical that leads to the formation of 2-Butynedinitrile and hydrogen.
Four different computational methods were benchmarked in order to check if even cheap methods can get accurate results for this system.
A comparison to Petrie and Osamura’s results was also done.
\indent
The paper is organized as follows: in Sec.~2, the methods and the construction of the potential energy surface are outlined. Preliminary results are reported in Sec.~3 and in Sec.~4 concluding remarks are given.

\section{Methods}

The Potential Energy Surface (PES) of the system was investigated through the optimization of the most stable stationary points.
Following a well established computational scheme~\cite{falcinelli2016stereoselectivity,leonori2009crossed,bartolomei2008intermolecular,de2011proton,leonori2009observation,de2007ssoh,rosi2012theoretical}, we optimized the geometry of the stationary points, both minima and saddle points, using a less expensive method with respect to the one employed in order to get accurate energies. 
Calculations for geometries were performed in order to benchmark three methods: density functional theory, with the Becke-3-parameter exchange and Lee-Yang-Parr correlation (B3LYP)~\cite{becke1993density,stephens1994ab}, Unrestricted-Hartree-Fock (UHF)~\cite{roothaan1951new,pople1954self}, and post Hartree-Fock B2PLYP~\cite{grimme2006semiempirical} combined or not with Grimme’s D3BJ dispersion~\cite{grimme2011effect,goerigk2011efficient}. 
All methods were used in conjunction with the correlation consistent valence polarized basis set aug-cc-pVTZ~\cite{dunning1989gaussian}. In each level of theory, a vibrational frequency analysis was done to confirm the nature of the stationary points: a minimum in the absence of imaginary frequencies and a saddle point if one and only one frequency is imaginary. The assignment of the saddle points was performed using intrinsic reaction coordinate (IRC) calculations~\cite{gonzalez1989improved,gonzalez1990reaction}. Then for each stationary point for all methods, the energy was computed with coupled cluster including single and double excitations and a perturbative estimate of connected triples (CCSD(T))~\cite{bartlett1981many,raghavachari1989fifth,olsen1996full}.
CCSD(T) is a more accurate method than the ones used for the optimizations, but prohibitive computational costs restricts the use of the method in this particular system to fixed geometry calculations. 
Finally, zero-point energy correction obtained through the frequency calculations were added to energies obtained from all methods to correct them to 0 K. 
All calculations were performed using the Gaussian 09 code~\cite{frisch2009gaussian}. 

\section{Results and discussion}

\begin{figure}[t]
\centering
\includegraphics[scale=0.25]{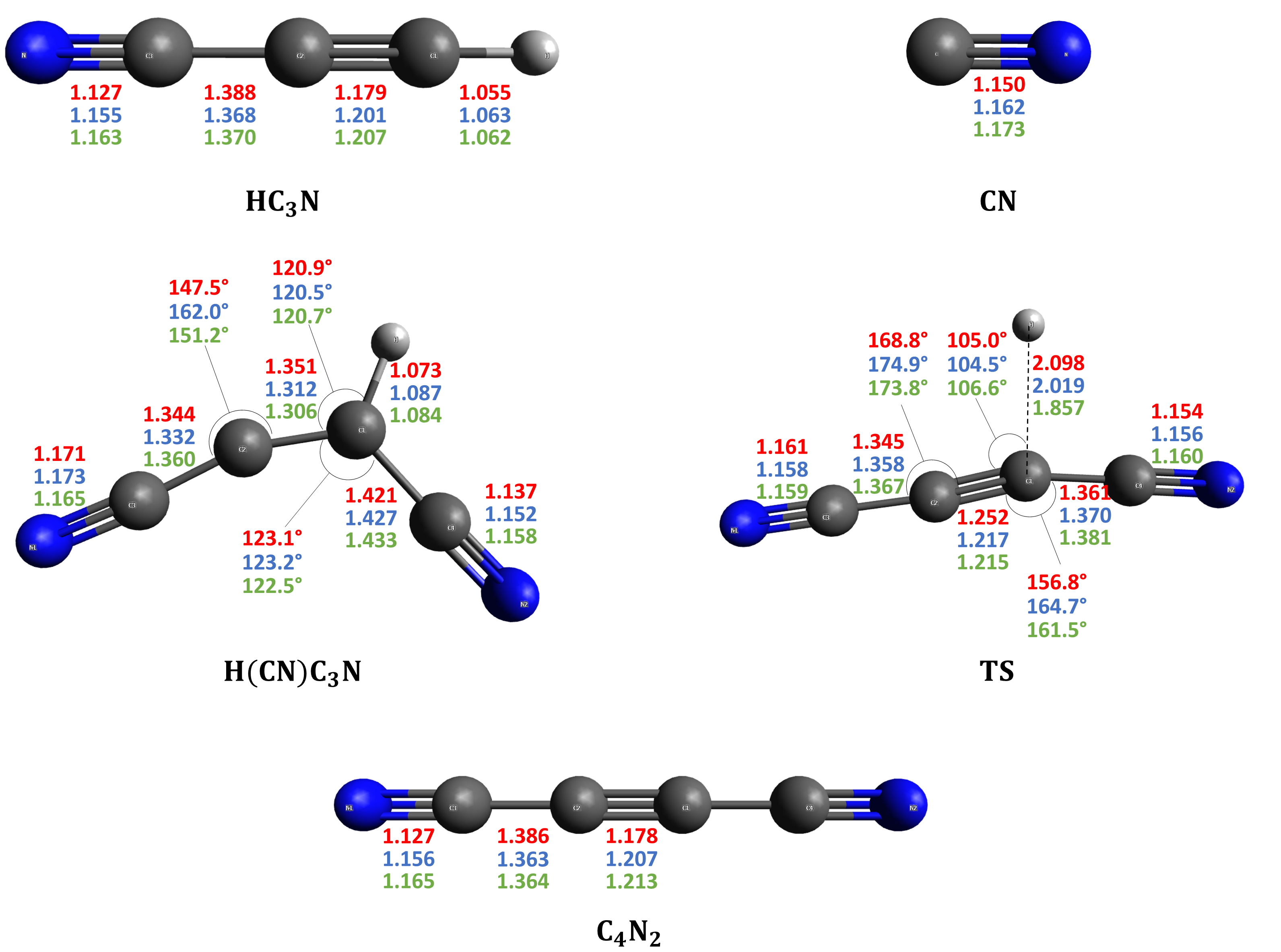}
\caption{Optimized geometries of the stationary points along the minimum energy path leading cyanoacetylene and cyano radical to 2-Butynedinitrile and hydrogen. 
Bond lengths are shown in \si{\angstrom} and bond angles are displayed in degrees. UHF values in red, B3LYP values in blue, B2PLYP values in green.}
\label {fig1}
\end{figure}

The calculation of the different electronic structures shows that the attack of cyano radical on the cyanoacetylene is an energetically favorable process that leads to the formation of an adduct intermediate.
Figure~\ref{fig1} gather the geometries at the minimum energy path, starting from the HC$_{3}$N and CN reactants and leading to 2-Butynedinitrile and hydrogen. These geometries were optimized at UHF/aug-cc-pVTZ, B3LYP/aug-cc-pVTZ, B2PLYP/aug-cc-pVTZ and B2PLYPD3/aug-cc-pVTZ levels. The reported interatomic distances are in red for UHF, blue for B3LYP and green for both B2PLYP and B2PYLP with Grimme’s D3 dispersion, since no difference in their geometries was recorded. The geometries obtained with the methods above are very similar. The differences between bond lengths and angles are small, with the exception of the transition state geometries.
In the saddle point structure, B2PLYP seems to provide a more reasonable distance for weak interaction as the one between carbon and hydrogen: it is 0.162 \si{\angstrom} shorter than the distance obtained with B3LYP and 0.241 \si{\angstrom} shorter than the one proposed by UHF.
B2PLYP is a double-hybrid density functional that combines Becke exchange and Lee, Young and Parr correlation with  Hartree-Fock exchange and a perturbative second-order correlation obtained from Kohn-Sham orbitals~\cite{grimme2006semiempirical}. According to the author of the method, B2PLYP reports very good results for transition state geometries.

\begin{table}[t]
\centering
\caption{Energies (kJ.mol$^{-1}$, 0 K) of the different geometries relative to the reactants. The energies are computed at UHF/aug-cc-pVTZ, B3LYP/aug-cc-pVTZ, B2PLYP/aug-cc-pVTZ, and B2PLYPD3/aug-cc-pVTZ levels of theory. In parentheses are the values for the same geometry computed at CCSD(T)/aug-cc-pVTZ level of theory.} \label{tab1}
\includegraphics[scale=0.4]{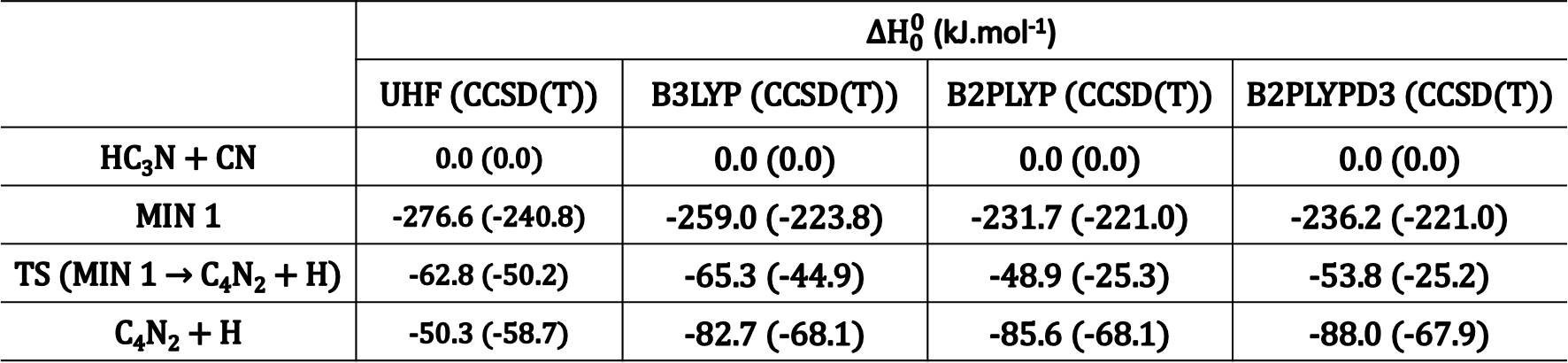}
\end{table}

Most of the geometries might be similar, but this does not necessarily means that all energies also are. 
Table~\ref{tab1} gathers the relative energies of all geometries computed with every method. 
The reactants were taken as the reference for the energy.
Within the B2PLYP method, the account of Grimme’s D3 dispersion changes slightly the energy value, even for an identical geometry.
As every method returns different energy values, a direct comparison is not pertinent.
However, at least the trend in the evolution of the energies can be compared. 
First, it can be observed that all stationary points are below the energy of the reactants in all the methods that have been employed.
Second, in all methods the adduct intermediate MIN 1 is the most energetically stable structure.
In addition, a barrier between the adduct and the products is well characterized for B3LYP and both B2PLYP methods, but it is not the case for the UHF method.
Though the saddle point was identified with UHF, the energy of the products returned a higher value.
 It will be shown, however, that when a single-point energy calculation with these same geometries optimized with UHF is done with the CCSD(T)/aug-cc-pVTZ method, the energy of the products is below the energy of the saddle-point.

In order to compare the geometries optimized with all those methods, single-point CCSD(T) calculations were also performed.
The results are reported again in Table~\ref{tab1}, in parenthesis.
Firstly, it can be observed that the energies of all stationary points, i.e. minima and transition state, are below the energy of the reactants in all the methods used.
Secondly, as B2PLYP and B2PLYPD3 optimized geometries were mostly identical, the energies computed with CCSD(T)/aug-cc-pVTZ were also the same.
Thirdly, it can be noticed that the energies of the B3LYP optimized geometries are very close to the ones of B2PLYP methods, with the exception of the transition state geometry: for this saddle point, energy difference is around 19.6 kJ.mol$^{-1}$.
At last, the energies of the UHF optimized geometries are the most different from the others methods. In particular, the energies of the intermediate and transition state gave the lowest energy at CCSD(T) level. In contrast, the energy of the products was the highest one using the UHF geometry.
Nevertheless, the energy barrier between the intermediate and the products is now well characterized for all methods.

It is also interesting to look at the barrier height values obtained for each method.
Table~\ref{tab2} reports the energy changes and barrier heights, computed at the same levels of theory, for the process leading to 2-Butynedinitrile and hydrogen.
The interaction of HC$_{3}$N and CN gives rise to the adduct MIN 1 (or H(CN)C$_{3}$N), more stable than the reactants in all levels of theory. This adduct evolves, through a barrier leading to the transition state, to the products of the reactive channel C$_{4}$N$_{2}$ and radical H. In all levels of theory, the products are less stable than the adduct.
Making a comparison between methods, UHF estimates the largest enthalpy changes and barrier height.  UHF is followed by B3LYP, B2PLYPD3 and B2PLYP in this order.
In relation to CCSD(T), a higher-level method, UHF and B3LYP energies are systematically overestimated. B2PLYP and B2PLYPD3 overestimates the CCSD(T) in the enthalpy variation attributed to the formation of the adduct, but underestimate both the barrier and enthalpy change for the reaction that leads the intermediate to the product.

\begin{table}[t]
\centering
\caption{Enthalpy changes and barrier height (kJ.mol$^{-1}$, 0 K) computed at UHF/aug-cc-pVTZ, B3LYP/aug-cc-pVTZ, B2PLYP/aug-cc-pVTZ, and B2PLYPD3/aug-cc-pVTZ levels of theory. In parentheses are the values for the same geometry computed at CCSD(T)/aug-cc-pVTZ level of theory.} \label{tab2}
\includegraphics[scale=0.31]{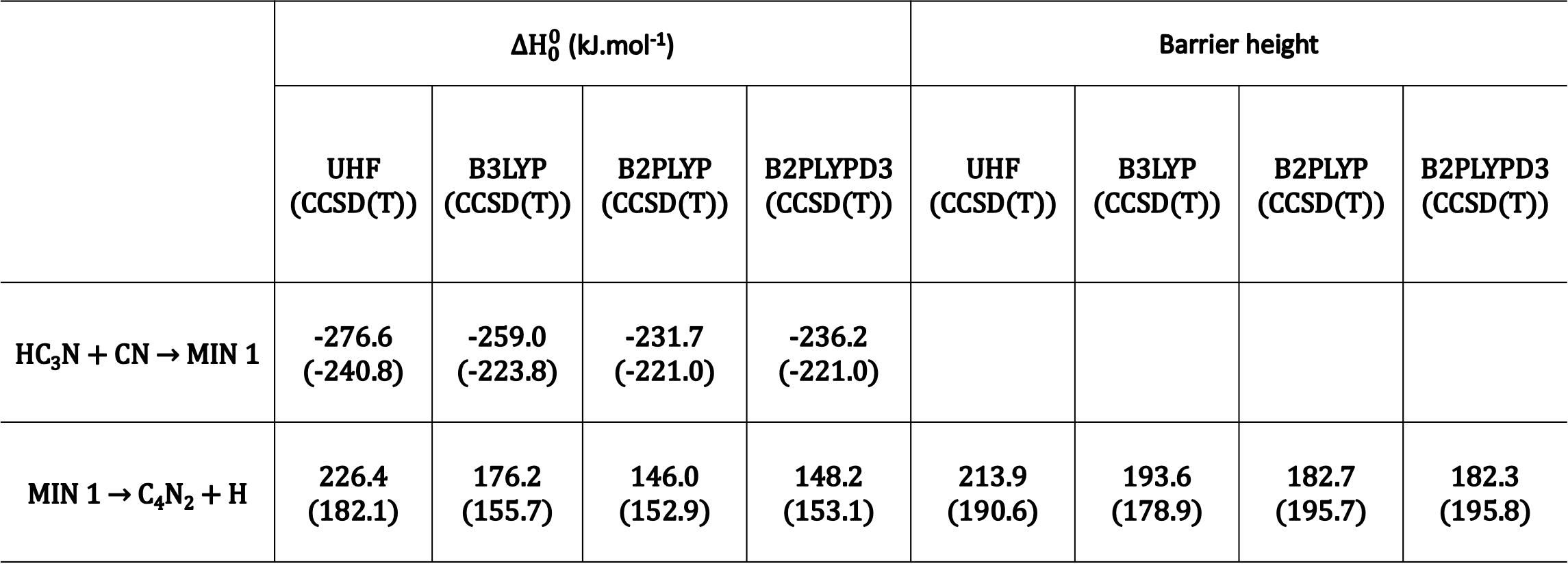}
\end{table}

While in this work the aug-cc-pVTZ basis set was employed for every method, Petrie and Osamura~\cite{petrie2004nccn} carried geometry optimizations at B3LYP level in conjunction with the 6-311G** basis-set and computed the single-point energies at CCSD(T) level in conjunction with aug-cc-pVDZ basis-set.
In respect to the geometries obtained, only the transition state and the intermediate geometries were published in their paper. The bond distance and angle values are similar to the ones in the B3LYP/aug-cc-pVTZ optimized geometries. The difference is smaller than when the B3LYP geometries were compared to the ones obtained with other methods.
The energies computed at CCSD(T)/aug-cc-pVDZ level are listed in Table 2 of that same paper.
Since the authors provided the values of total energy for all stationary points, it was possible to calculate the energy of each point in relation to HC$_{3}$N + CN.
The energy at 0 K is -231.1 kJ.mol$^{-1}$ for the adduct intermediate, -39.5 kJ.mol$^{-1}$ for the transition state and -58.6 kJ.mol$^{-1}$ for the products.
The energies of the stationary points are once again below the energy of the reactants. 
CCSD(T)/aug-cc-pVDZ  provides here a lower energy value for the adduct intermediate and a higher energy value for the transition state and the products than CCSD(T)/aug-cc-pVTZ.
At 191.6 kJ.mol$^{-1}$, the height of the barrier between the adduct intermediate and the products is larger in comparison to CCSD(T)/aug-cc-pVTZ.
On the other hand, at 172.5 kJ.mol$^{-1}$ the enthalpy change between the adduct and the product is very close to the one showing on Table~\ref{tab2}.

\section{Conclusions}
As far as optimized geometries for stationary points are concerned, the UHF method seems to be inadequate, while DFT methods seem to be more reliable. In particular, B3LYP functional seems to work well for minima, while better functionals like the B2PLYP seem to be necessary for transition state geometries when van der Waals interactions are present. As far as energies are concerned, however, only more correlated methods like CCSD(T) seems to provide accurate results.\\
\indent
A more general conclusion concerns the reaction mechanism suggested by our calculations, which is also in line with the previous determination by Petrie and Osamura. Similarly to the case of other reactions involving CN and other species holding a triple C$-$C bond (such as ethyne, propyne and 2-butyne \cite{Huang1999jcp,Balucani1999jcp,Huang2000jcp}), the CN radical interacts with the electron density of triple bond to form an addition intermediate without an activation energy. This is in line with the large value of the rate coefficient derived for the reactions also at very low temperature \cite{cheikh2013low} and makes this process a feasible route of dicyano-acetylene (2-Butynedinitrile) even under the harsh conditions of the interstellar medium. \\ 
\indent
After completing the derivation of the potential energy surface for the title reaction, we will run kinetic calculations to derive the rate coefficient and product branching ratio. The calculated rate coefficient will be compared with the experimental values derived by Cheikh et al. \cite{cheikh2013low} while the reaction mechanism and product branching ratio will be compared with those inferred by the crossed molecular beam experiments which are now in progress in our laboratory. A thorough characterization of the title reaction will allow us to establish its role in the nitrogen chemistry of the interstellar medium. To be noted that both cyano- and dicyano-acetylene have a strong prebiotic potential \cite{balucani2012elementary}.

\section{Acknowledgements}

This project has received funding from the European Union’s Horizon 2020 research and innovation programme under the Marie Skłodowska-Curie grant agreement No 811312 for the project "Astro-Chemical Origins” (ACO).
E.V.F.A. thanks the Dipartimento di Ingegneria Civile ed Ambientale of University of Perugia for  allocated computing time. 
N.F.L. thanks Perugia University for financial support through the AMIS project (“Dipartimenti di Eccellenza-2018–2022”),  also thanks the Dipartimento di Chimica, Biologia e Biotecnologie for funding under the program Fondo Ricerca di Base 2017.
M.R. acknowledges the project “Indagini teoriche e sperimentali sulla reattività di sistemi di interesse astrochimico” funded with Fondo Ricerca di Base 2018 of the University of Perugia.

\bibliography{evfa_iccsa2020}{}

\bibliographystyle{splncs} 
\end{document}